**Expanding the extreme-κ dielectric materials space through physics-validated generative reasoning**

**Authors:** Hossain Hridoy[1], Tahiya Chowdhury[2], Md Shafayat Hossain[3,4,5]*

**Affiliations:**

[1]Department of Chemical Engineering, Bangladesh University of Engineering and Technology, Dhaka, Bangladesh.

[2]Department of Computer Science, Colby College, Waterville, ME, USA.

[3]Department of Materials Science and Engineering, University of California, Los Angeles, CA, USA.

[4]California NanoSystems Institute, University of California, Los Angeles, CA, USA.

[5]Center for Quantum Science and Engineering, University of California, Los Angeles, CA, USA.

*Correspondence to: mshossain@ucla.edu

## Abstract

The most technologically consequential materials are often the rarest: they occupy narrow regions of chemical space, obey competing physical constraints, and appear only sparsely in existing databases. High-κ dielectrics, high-$T_c$ superconductors, and ferromagnetic insulators are to name a few. This scarcity fundamentally limits today's data-driven materials discovery, where machine-learning models excel at interpolation but struggle to generate genuinely new candidates. Here, we introduce *DielecMIND*, an artificial intelligence framework that reframes materials discovery as a reasoning-driven exploration instead of a database-screening problem. Using high-κ dielectrics as a data-scarce and technologically stringent test case, *DielecMIND* combines large-language-model hypothesis generation for the first time with physics validated first-principles calculation to navigate chemical space beyond known compounds. Prior to our work, only 14 experimentally or computationally validated materials with κ > 150 were known. Our framework discovers and validates 5 new such compounds, expanding this rare-materials class by a remarkable ≃35% in a single study. Among them, we find that $Ba_2TiHfO_6$ exhibits a dielectric constant of 637, minimal loss at low optical frequencies, and stability up to 800 K. Beyond dielectrics, this work demonstrates a new paradigm for artificial-intelligence-guided discovery: one that generates a small number of physically grounded, experimentally plausible candidates yet measurably expands sparsely populated functional materials spaces. Thus, *DielecMIND* points toward a general strategy for discovering rare, high-impact functional materials where data scarcity has long constrained progress.

## 1. Introduction

A central challenge in modern materials science is not the lack of materials database, but the fact that only a handful of known compounds often represent the most technologically consequential materials classes. At the same time, data-driven discovery methods are optimized for regimes with thousands to millions of labeled examples. This tension is especially acute for functional materials that must satisfy tightly coupled, multi-objective constraints, where extrapolation beyond known chemistries is essential, but training data are intrinsically sparse. Here, we introduce for the first time a large-language-model (LLM)–guided discovery framework designed explicitly for such data-scarce regimes, and we use high-permittivity ($\kappa$) dielectric materials as a representative and technologically urgent testbed to demonstrate its capabilities. Despite decades of dielectric research, there are only 14 known compounds found in the Materials Project database with static dielectric constants exceeding κ ≈ 150 prior to our work, underscoring how narrowly confined the high-$\kappa$ design space remains. Each additional verified material in this regime represents a statistically meaningful expansion of known chemical space. Here, our framework adds 5 new members to this population, expanding the known high- $\kappa$ landscape by ~35%, a scale of progress that would traditionally require many years of iterative theory–experiment cycles.

The significance of high- $\kappa$ dielectric materials cannot be overstated. The continued scaling of modern electronics increasingly hinges not on lithography alone, but on the availability of materials that can sustain functionality as device dimensions approach fundamental physical limits. This challenge is particularly acute in dynamic random-access memory (DRAM) and advanced logic technologies, where capacitors and gate stacks must reliably store charge as footprints shrink toward late–10 nm technology nodes [1,2]. In DRAM, a single transistor–capacitor pair encodes information via stored charge; as lateral dimensions contract, maintaining sufficient capacitance forces dielectric layers into thickness regimes where leakage, variability, and reliability dominate performance. As a result, dielectric choice and integration have emerged as central bottlenecks to continued scaling.

Historically, $SiO_2$ served as the dielectric workhorse of microelectronics due to its large band gap ($E_g \approx 8.9$ eV) and excellent interfacial quality. However, its modest dielectric constant ($\varepsilon \approx 3.9$) required aggressive thickness reduction to achieve the target capacitance, ultimately leading to an unavoidable physical barrier: as thickness decreases, quantum tunneling and defect-assisted leakage currents rise sharply, degrading retention and reliability [3–5]. This limitation motivated the adoption of high-permittivity (high-κ) dielectrics such as $BaTiO_3$ and $SrTiO_3$, which enable higher capacitance without pushing thickness into the most leakage-prone regime. Beyond integrated circuits, high-κ oxides are foundational to multilayer ceramic capacitors (MLCCs), now ubiquitous in modern electronics for decoupling, DC blocking, filtering, and signal conditioning [6–8]. The rapid growth of MLCC usage further underscores the importance of discovering materials that simultaneously combine high permittivity, wide band gap, low loss, and practical stability.

Such materials remain rare because the problem is intrinsically multi-objective and rooted in materials physics. The dielectric response comprises an electronic contribution ($\varepsilon_\infty$), arising from electron cloud displacement, and an ionic contribution ($\varepsilon_0$), associated with lattice polar modes, such that $\varepsilon = \varepsilon_\infty + \varepsilon_0$ [9]. Large dielectric constants often originate from soft lattice dynamics and strong ionic polarizability. Yet, extensive empirical and theoretical studies have revealed a persistent trade-off between dielectric constant and band gap across oxide families [10]. Practical high-κ dielectrics therefore demand balance rather than single-property optimization, making brute-force exploration inefficient.

Traditionally, identifying such balanced candidates has relied on tightly coupled theory–experiment cycles involving synthesis, characterization, and physics-based modeling. First-principles approaches such as density functional theory (DFT), density functional perturbation theory (DFPT) and ab initio molecular dynamics (AIMD) provide quantitative rigor but are computationally intensive, rendering exhaustive chemical-space searches impractical without strong candidate triage [11]. Although the Materials Genome Initiative catalyzed the creation of large materials-property databases and data-driven discovery paradigms [12], machine-learning approaches remain constrained in sparse-label regimes. Existing ML models have demonstrated success in specific dielectric domains, including polymers [13], $ABX_3$ perovskites [14], ceramic oxides [15], organic solvents [16], and graph neural network–based prediction of inorganic dielectric constants [17]. However, these methods preferentially rank candidates near the training manifold and struggle to propose genuinely novel chemistries, particularly when dielectric labels are unevenly distributed and scarce [18].

Large language models introduce a fundamentally different capability. Rather than learning solely from numerical correlations, LLMs can encode domain knowledge expressed in natural language, reason across concepts, and orchestrate multi-step workflows. Recent studies have shown promise in applying LLMs to materials-related tasks such as battery cathode identification [19], polymer solubility prediction [20], crystal geometry inference [21], and synthesis planning [22]. Yet, when used in isolation, LLMs are unreliable predictors of quantitative materials properties; their autoregressive reasoning can produce incomplete or inconsistent logic without grounding in physical constraints and validated computation [23]. To overcome such constraints, we propose that LLMs with domain-encoded knowledge can be a potential ground for discovering functional materials, particularly in data-scarce regimes like high-κ dielectrics.

Here, we introduce *DielecMIND*, a materials discovery framework that couples an LLM with domain-embedded knowledge and closes the loop through first-principles validation using DFT/DFPT and AIMD. Rather than replacing physics-based computation, *DielecMIND* uses LLM-guided reasoning to navigate sparse, high-impact regions of chemical space while enforcing quantitative rigor through explicit verification. Applying this framework, we generate a prediction pool of 120 materials, from which we identify 5 previously unknown high-κ dielectric candidates, expanding the known high-κ (with κ exceeding 78) landscape from 14 to 19. The top-ranked material, $Ba_2TiHfO_6$, exhibits a dielectric constant of 637.01 and a figure-of-merit of 1547.93, placing it among the extreme high-κ tail of the Materials Project database, where only ~0.05% of dielectrics exhibit κ values above 600. Detailed analysis further reveals near-zero dielectric loss tangent at low photon energies, while thermodynamic stability up to 800 K is confirmed by AIMD.

More broadly, this work demonstrates for the first time a generalizable paradigm for materials discovery in regimes where data are scarce but functional demands are stringent. By integrating human-readable domain knowledge, prompt-engineered reasoning, and first-principles verification, this framework opens a new frontier for AI-driven materials discovery beyond interpolation—one that is especially suited to the rare, high-impact materials that ultimately define technological progress.

## 2. Results and discussion

### 2.1. Framework architecture

*DielecMIND* is designed to address a central tension in dielectric materials discovery: the search space is vast and weakly labeled, while the underlying physics that governs high dielectric response is subtle, multi-objective, and highly structured. Rather than treating discovery as a single-shot prediction task, *DielecMIND* implements an iterative workflow in which candidate selection, physical reasoning, and novelty enforcement are explicitly separated and coordinated. The overall architecture of the workflow, illustrated in Fig. 1, consists of two complementary phases that differ in how domain knowledge and physical constraints are injected into the generation loop. An LLM is prompted to propose candidate dielectric materials in a structured JSON format, after which a specialized tool validates the novelty of these hypotheses. Each iteration yields a single candidate, and the loop is repeated until a curated set of 60 valid high-κ dielectric candidates is produced. The distinction between these two phases lies in their strategy: Phase I prioritizes broad, novelty-driven exploration, whereas Phase II emphasizes dielectric material design-guided refinement and interpretability.

**Phase I: Zero-shot, domain-constrained exploration.** Phase I is designed to rapidly explore chemical space while enforcing basic feasibility and novelty constraints. This phase relies on two components: domain-constrained reasoning engine, and search tool. The domain-constrained reasoning engine performs broad, zero-shot exploration over a large curated dataset (see section 4.1 for dataset curation) by enforcing high-level expert knowledge of desirable dielectric characteristics through explicit prompting rules for the LLM. In particular, candidate materials are required to satisfy two target conditions: (i) a wide electronic band gap exceeding 2–3 eV and (ii) a static dielectric constant at the colossal scale.

Using zero-shot prompting—i.e., generative inference without in-prompt demonstrations, guided by domain-encoded prompting rules—the domain-constrained engine instructs the LLM to propose new dielectric compositions and structures by modifying or recombining reference materials drawn from large curated dataset. Importantly, the engine explicitly enforces the generation of candidates that do not already exist in the MP database, thereby biasing the search toward unexplored chemical regimes. Nevertheless, due to the possibility of hallucinated or redundant outputs, a specialized search tool is introduced as a gatekeeper. This tool queries the MP database via its API to verify whether a proposed candidate is already cataloged and rejects any duplicates.

Because LLM outputs may violate structural or compositional constraints, the generation loop is repeated until valid candidates satisfying all criteria are obtained. Accepted candidates are stored in a standardized JSON schema for downstream processing. Details of the prompt design for the domain-constrained reasoning engine are provided in Appendix A.1.

**Phase II: Chain-of-thought (CoT), dielectric material design-guided reasoning.** While Phase I emphasizes breadth, Phase II is explicitly designed to embed physical insight into the generation process. This phase addresses a known limitation of LLMs: performance degrades when reasoning over large, unstructured input dataset [24]. To mitigate this, *DielecMIND* introduces a clustering module that compresses large curated dataset into physically meaningful subsets suitable for reasoning (Fig. 2).

In each iteration, the clustering module selects a subset of four chemically related materials as reference structures. Subset selection is based on two criteria: (i) the total number of atoms common across all four formulas (primary score) and (ii) the mean pairwise Jaccard similarity of their elemental compositions (secondary score). A greedy search evaluates all candidates as potential seeds and selects the subset using lexicographic optimization, first prioritizing maximal shared elemental composition and then the mean pairwise Jaccard similarity. This strategy ensures that the LLM reasons over compact, chemically coherent exemplars rather than diffuse dataset. The implementation of this module is described in Appendix B.

With the reference subset defined, a dielectric design-guided reasoning engine is invoked using CoT prompting. This engine focuses the generative process toward maximizing dielectric response through structured design rules and screening criteria. Specifically, it leverages lattice-dynamical mechanisms known to enhance dielectric response. The ionic contribution to the static dielectric constant is governed by the Born effective charges $Z_m^*$ and phonon frequencies $\omega_m$,

$$\varepsilon_0 \propto \sum_m \frac{(Z_m^*)^2}{\omega_m^2},$$

where second-order Jahn–Teller distortions associated with $d^0$B-site cations in perovskites generate anomalously large $Z_m^*$, enhancing $\varepsilon_0$[25–29]. Additional mechanisms include the incorporation of stereoactive lone-pair cations at the A-site, which reduce local symmetry and increase polarizability [30,31], and aliovalent A- or B-site doping to induce defect dipoles that further enhance dielectric response [32,33].

Based on these principles, the dielectric design-guided reasoning engine proposes three candidate formulas per iteration and filters them according to quantitative constraints, including band gap $E_g > 2.4$eV, ionic dielectric constant $\varepsilon_0 \geq 150–300$, and figure-of-merit exceeding 300. The most promising candidate is then selected, and a corresponding POSCAR structure is generated in JSON format (Table S3 in Appendix H). Prompt details are provided in Appendix A.2. As in Phase I, a final MP database check is performed by the search tool to enforce novelty before acceptance.

Through the complementary operation of these two phases, *DielecMIND* balances exploration with physical rigor, enabling the discovery of high-κ dielectric materials that are both novel and quantitatively grounded.

## 2.2. Performance evaluation

We evaluate *DielecMIND* through a series of holistic studies: (i) how well its two phases internalize dielectric material design when asked to predict properties, (ii) how efficiently the best workflow generates actionable candidates for first-principles verification, and (iii) how reasoning overhead affects the generative results of the framework.

**Dielectric performance across two phases.** The performance evaluation begins by benchmarking the two-phase workflows: (i) the Phase I pipeline, in which the large curated dataset is processed using the domain-constrained reasoning engine, and (ii) the Phase II pipeline, which performs clustering-based dataset curation prior to applying the dielectric design-guided reasoning engine. Using DFT and DFPT datasets from the literature, we benchmark predictions of the ionic dielectric constant ($\varepsilon_0$), electronic dielectric constant ($\varepsilon_\infty$), total dielectric constant (ε), and band gap ($E_g$) [17,34]. For a controlled comparison, we switch both reasoning engines from structure generation to property prediction. Each benchmark formula is passed into the reasoning engine via an input function, and the model is queried only for the corresponding dielectric properties. Phase I produces a direct zero-shot estimate based on its large curated dataset knowledge, whereas Phase II first curates selected subsets using the clustering module based on the benchmark formula and then applies the structured COT template that explicitly invokes dielectric material design-embedded reasoning before outputting a prediction.

We quantify error using standard regression metrics (root mean square error, RMSE; coefficient of determination, $R^2$), summarized in Fig. 3. Phase I shows limited accuracy for dielectric constants, with $R^2$ values of 0.478 ($\varepsilon_\infty$), 0.317 ($\varepsilon_0$), and 0.318 (ε), and RMSE values of 0.089 ($\varepsilon_\infty$), 0.277 ($\varepsilon_0$), and 0.27 (ε).

Phase II substantially improves dielectric prediction, reaching $R^2$ values of 0.604, 0.677, and 0.685 and reducing RMSE to 0.062, 0.193, and 0.184 for $\varepsilon_\infty$, $\varepsilon_0$, and $\varepsilon$, respectively. For band gap prediction, Phase I exhibits an $R^2$ of 0.576 with RMSE of 0.049; Phase II maintains predictions close to the dashed parity line in Fig. 3d, with an $R^2$ of 0.762 and RMSE of 0.041. Taken together, these benchmarks show that zero-shot prompting struggles to capture the material design required to reason reliably about high-κ behavior, whereas dielectric design-embedded COT improves the dielectric-property predictions that matter most for screening and prioritization.

**Evaluation across different models.** We compare the best-performing phase of *DielecMIND* against prior hypothesis-generation approaches for high-κ dielectrics, summarized in Table 1. We use precision, defined as the fraction of predicted candidates that satisfy the target figure-of-merit (FOM, $\Phi_M$) after DFT/DFPT verification, divided by the total number of predicted candidates. Under this metric, Phase II achieves a precision of 8.3%, exceeding an artificial neural network (ANN) ensemble model (5.8%) and an equivariant graph neural network (EGNN) model (5%) [17,35]. A joint EGNN plus machine learning potential (MLP) approach reported by Takigawa et al. achieves a higher precision of 15.8% [34]. However, these reported models identify candidates drawn from the MP database whose dielectric constants have not yet been computed, whereas our framework is designed to generate candidate materials that are not present in the MP database. In this sense, *DielecMIND* is not only a filter but also a generator that can expand the reachable chemical space.

**Baseline evaluations.** For a systematic comparison, we assessed *DielecMIND* against a chemically constrained elemental-substitution baseline. To maintain logical consistency with *DielecMIND*, the baseline was built by applying chemically meaningful substitute pathways in the curated dataset, such as i. lone pair (Pb, Sn; Bi, Sb), alkaline earth (Ba, Sr, Ca), and rare-earth (La, Nd, Sm, Gd, Y) A-site cations; ii. lattice polarizable $d^0$ (Ti, Zr, Hf; Nb, Ta) and trivalent (Sc, In) B-site cations. Detailed algorithm of the baseline implementation is available in Appendix C. Under DFT/DFPT validation, only two materials from baseline, $Zr_2Ga_4O_{10}$ ($\varepsilon_0$ = 33.86, $\varepsilon_\infty$ = 4.89, $\varepsilon$ = 38.75, $E_g$ = 3.10 eV) and $Ca_2Hf_3O_8$ ($\varepsilon_0$ = 11.94, $\varepsilon_\infty$ = 3.70, $\varepsilon$ = 15.64, $E_g$ = 4.59 eV), show dielectric behavior. Detailed theoretical analysis of these two is provided in Fig. S1 of Appendix D. Therefore, the baseline yields a lower precision rate of around 3.3% over a FOM value of 70, which is much lower than the precision and FOM values of our *DielecMIND* framework. This result justifies the strong potential of *DielecMIND* as a predictive framework for the discovery of high-κ dielectric materials.

**API cost and inference latency.** We quantify the costs and latency, because a discovery workflow is only useful if it can iterate fast enough to support effective triage. We measure average end-to-end wall-clock inference latency and token usage per sample for each phase. Phase I is faster, averaging 105.3 seconds per sample, at a cost of \$0.147 for 66.25k tokens (input plus output), but this speed comes at the expense of reduced prediction quality. Phase II is slower, averaging 163.1 seconds per sample and costing \$0.249 for 67.15k tokens (input plus output), reflecting the additional reasoning steps introduced by COT prompting. This latency overhead is the direct price of dielectric design-guided reasoning, and the benchmark results in Fig. 3 and Table 1 show when that price is justified for high-κ discovery.

## 2.3. DFT/DFPT verification

To establish the physical validity of the candidates proposed by *DielecMIND*, we performed systematic first-principles verification of all generated structures using DFT and DFPT. This step serves as the quantitative anchor of the framework, translating language-model–generated hypotheses into rigorously validated dielectric properties. Among the 60 structures generated in Phase I, four candidates exhibited dielectric behavior upon DFT/DFPT evaluation. However, none of the materials satisfied the criteria for a high-κ dielectric, although $BaBiScZrO_6$ reached a peak FOM of 94.35. In contrast, Phase II yielded a substantially higher success rate: five materials exhibited FOM values exceeding 350, highlighting the impact of dielectric design-guided reasoning and structured hypothesis generation. The full dielectric properties of all nine verified candidates are summarized in Table S2 in Appendix H.

Figure 4a situates these newly discovered materials within the extreme tail of the dielectric distribution in Materials Project, a regime occupied by fewer than ~0.38% of known dielectrics with $\kappa > 150$ and only ~0.05% with $\kappa > 600$. The candidates follow the established Pareto front linking dielectric constant (ε) and band gap ($E_g$) [36], underscoring the intrinsic trade-off between polarizability and electronic insulation. Notably, most Phase I candidates occupy regions with larger band gaps but more modest dielectric response, whereas Phase II candidates shift toward the extreme high-κ regime, reflecting the framework's ability to navigate this trade-off more aggressively while retaining stability. Structural motifs for all eleven verified candidates, including two from baseline, are shown in Fig. 4b. Five materials among those are identified as high-κ dielectrics: $Ba_2TiHfO_6$ ($\varepsilon_0 = 631.3$, $\varepsilon_\infty = 5.71$, $\varepsilon = 637.01$, $E_g = 2.43$ eV), $PbSrTi_2O_6$ ($\varepsilon_0 = 243.11$, $\varepsilon_\infty = 6.33$, $\varepsilon = 249.44$, $E_g = 2.56$ eV), $BaPbTi_2O_6$ ($\varepsilon_0 = 170.89$, $\varepsilon_\infty = 5.92$, $\varepsilon = 176.81$, $E_g = 2.46$ eV), $BaSrCaTi_3O_9$ ($\varepsilon_0 = 151.8$, $\varepsilon_\infty = 5.53$, $\varepsilon = 157.33$, $E_g = 2.56$ eV), and $NaKTa_2O_6$ ($\varepsilon_0 = 146.73$, $\varepsilon_\infty = 5.02$, $\varepsilon = 151.75$, $E_g = 2.45$ eV). Prior to our work, only 14 materials with $\kappa > 150$ were known across the entire Materials Project database. The addition of these 5 new compounds therefore represents an expansion of approximately 35% of the known extreme-κ materials landscape.

Among these, $Ba_2TiHfO_6$ and $PbSrTi_2O_6$ stand out as the top-performing candidates. We, therefore, analyzed their structural, electronic, and dynamical properties in detail. Crystallographic analysis reveals that $Ba_2TiHfO_6$ crystallizes in a cubic lattice with space group *Fm-3m*, while $PbSrTi_2O_6$ adopts a tetragonal structure with space group *P4mm* (Figs. 5a and 5b). Their electronic structures, shown in Figs. 5c and 5d, further confirm their suitability as dielectric materials. $Ba_2TiHfO_6$ exhibits a direct band gap of 2.43 eV at the Γ point, whereas $PbSrTi_2O_6$ displays an indirect gap of 2.56 eV, with the valence band maximum at the M point and the conduction band minimum at Γ. As is well established, conventional DFT underestimates the band gaps of insulating oxides by approximately 40% [37], suggesting that the true experimental gaps are larger.

Beyond static properties, we evaluated frequency-dependent dielectric loss to assess functional performance. Figure 5e shows that both materials exhibit a near-zero dielectric loss tangent across low photon energies (approximately 0–5 eV), indicating minimal energy dissipation and strong insulating behavior in technologically relevant frequency ranges. Finally, thermal stability was assessed using *ab initio* molecular dynamics (AIMD). As shown in Fig. 5f, both materials remain structurally intact up to 800 K, with smooth temperature fluctuations and only minor atomic displacements (Fig. S2 in Appendix E). No bond breaking or large-scale structural distortions are observed, confirming robust thermodynamic stability. The phonon characteristics of $Ba_2TiHfO_6$ and $PbSrTi_2O_6$ are represented in Fig. S3a and b with Table S1 in Appendix F, respectively, showing no imaginary phonon frequencies. Since both of the materials are dynamically stable, their cation ordering should be synthesizable [43]. Taken together, these results demonstrate that *DielecMIND* does not merely generate speculative candidates but produces quantitatively validated, thermodynamically stable, and functionally competitive high-κ dielectric materials, including compounds that reside in the extreme tail of known dielectric performance.

## 2.4. Failure case analysis

To understand *DielecMIND*'s limitations and identify pathways for systematic improvement, we analyzed all failure-case materials generated across both phases. Three dominant failure modes emerged. The most frequent arose from imaginary polar phonon modes at the Γ point, which led to divergent or ill-defined ionic dielectric constants because $\varepsilon_0$ scales inversely with the square of the phonon frequency. A second class of failures involved electronic instability, in which candidates exhibited metallic or semimetallic band structures and therefore failed basic insulating criteria. Finally, a subset of candidates displayed geometrically unstable or unphysical structures, including broken coordination environments or distorted local bonding, despite satisfying compositional constraints at the hypothesis stage.

Notably, these failures persist even under stringent DFPT convergence settings, including dense q-point sampling, indicating that they are intrinsic to the generated structural motifs rather than numerical artifacts. A clear pattern is that *DielecMIND* preferentially proposes high-symmetry crystal structures, which prior studies have shown to be statistically more prone to dynamical instabilities due to soft polar modes and symmetry-protected lattice instabilities [38]. This observation suggests that future iterations of the framework should explicitly bias generation toward lower-symmetry or symmetry-broken structures, where polar distortions can stabilize large dielectric responses without triggering lattice instability. Notably, these failure modes do not undermine the validity of the discovered high-κ candidates; instead, they delineate a well-defined frontier for improving prompt design and dielectric material design-aware constraints, highlighting how AI frameworks can iteratively learn not only from successes but also from structured failure.

# 3. Conclusion

*DielecMIND* discovers 5 new high-κ dielectrics in a regime where only 14 were previously known, demonstrating that material design-aware reasoning can measurably expand rare functional materials classes. In materials spaces limited by scarcity rather than optimization, such stepwise increases in the known population represent a meaningful advance. More broadly, our work potentially reframes success in materials discovery for rare-property regimes. When progress depends on expanding an extremely small design space, impact can be best measured by the ability to access and validate new regions of chemical and structural space. *DielecMIND* exemplifies a discovery paradigm in which LLMs operate as material design-grounded scientific reasoners, navigating sparse and biased data landscapes by embedding domain knowledge, enforcing novelty, and closing the loop with first-principles validation. The resulting expansion of the extreme-dielectric landscape enables new directions in understanding ultra-high-permittivity structure–property relationships and assessing device-relevant stability. More generally, this framework is transferable to other material classes defined by rarity, offering a path toward accelerated discovery in which each new material has the potential to make transformative advances in science and technology.

## 4. Acknowledgments

We acknowledge illuminating discussions with Bruce Dunn.

## 5. Author Contributions

H.H. built the AI model in consultation with T.C. and M.S.H. H.H. performed the first-principles calculations in consultation with M.S.H. M.S.H. conceived and supervised the project. All authors contributed to the manuscript preparation and review.

## 6. Funding

M.S.H. acknowledges support from the Samueli Foundation (no specific grant number) and The UCLA Council on Research (no specific grant number).

## 7. Data Availability

All data supporting the findings of this study are available within the paper.

# 8. Methods

## 8.1. Dataset curation

The candidate space explored by *DielecMIND* is based on a curated, first-principles database to ensure physical realism from the outset. We constructed the input dataset from the MP database, version v2025-06-09, a comprehensive open-access repositories of DFT–computed materials properties [39]. To focus the search on technologically relevant dielectrics while maintaining broad chemical diversity, the database was filtered using two minimal criteria: a total dielectric constant $\varepsilon \geq 0.1$ and a band gap $E_g \geq 2$ eV. This filtering step yielded 3,646 candidate materials spanning a wide range of crystal systems, chemistries, and dielectric responses.

For each material, the dataset includes chemical formula, crystal system, space group, formation energy, energy above the convex hull, static ionic dielectric constant ($\varepsilon_0$), high-frequency electronic dielectric constant ($\varepsilon_\infty$), total dielectric constant ($\varepsilon$), band gap ($E_g$), and fully relaxed crystal structure. The resulting distribution of dielectric constants, shown in Fig. S4 of Appendix G, highlights the strong imbalance characteristic of dielectric datasets, namely, a dense population of modest-κ materials and a long, sparsely populated high-κ tail. This imbalance underscores the central challenge addressed in this work: how to prioritize genuinely novel, extreme dielectric candidates under conditions of sparse labels and skewed property distributions.

## 8.2. Large language model

At the core of the *DielecMIND* framework is GPT-5, a state-of-the-art LLM developed by OpenAI. The model was accessed through official API credentials and deployed as a hypothesis reasoning engine rather than as a direct numerical predictor. In contrast to conventional ML models trained explicitly on dielectric labels, GPT-5 is leveraged for its ability to encode and manipulate domain knowledge expressed in natural language, enabling structured hypothesis generation, constraint-aware reasoning, and iterative refinement within a prompt-engineered workflow. Crucially, the outputs of the LLM are continuously filtered and validated by first-principles calculations, ensuring that generative flexibility does not compromise physical plausibility. This hybrid design allows *DielecMIND* to explore beyond the immediate training manifold of existing datasets while remaining anchored to quantitative materials physics.

## 8.3. Computational details

All first-principles calculations used to validate *DielecMIND*-generated candidates were performed using the CASTEP code [40]. Electronic structure calculations were carried out within density functional theory using the generalized gradient approximation (GGA) with the Perdew–Burke–Ernzerhof (PBE) exchange–correlation functional [41]. Norm-conserving pseudopotentials were employed to describe the interaction between valence electrons and atomic cores, ensuring transferability across diverse chemical environments. Structural optimizations were conducted using the Broyden–Fletcher–Goldfarb–Shanno (BFGS) algorithm, which provides robust convergence for complex oxide geometries and is essential for obtaining accurate charge densities and electronic wave functions [42]. Pulay density mixing was applied to accelerate self-consistent field (SCF) convergence. The SCF electronic convergence criterion was stringently set to $5 \times 10^{-7}$ eV per atom. The plane-wave cutoff energies and k-point densities used throughout geometry relaxation and dielectric calculations ranged from 700 to 940 eV and 0.07 $Å^{-1}$, respectively. During relaxation, the convergence thresholds for Hellmann-Feynman force and ionic displacement were set to 0.01 eV/Å and $5 \times 10^{-4}$ Å, respectively. Furthermore, the energy difference per atom and maximum stress were fixed at $5 \times 10^{-6}$ eV and 0.02 GPa, respectively.

Dielectric properties were computed using DFPT, providing access to both the electronic ($\varepsilon_\infty$) and ionic ($\varepsilon_0$) contributions to the dielectric response. Recognizing the well-known tendency of standard DFT to underestimate band gaps in insulating materials [37], we employed the DFT+U approach where appropriate to improve the fidelity of the electronic structure. In selected cases, external strain was applied to explore its influence on electronic structure calculations.

Thermodynamic stability at finite temperature was assessed using AIMD simulations performed on $2 \times 2 \times 2$ supercells within the NVT ensemble. These simulations enable direct evaluation of lattice stability and structural integrity under thermal fluctuations, providing an essential complement to static zero-temperature screening.

## References

1. Kim, S. E., Sung, J. Y., Yun, Y., Jeon, B., Moon, S. M., Lee, H. B., Lee, C. H., Jung, H. J., Lee, J., & Lee, S. W. (2024). Atomic layer deposition of high-k and metal thin films for high-performance DRAM capacitors: A brief review. *Current Applied Physics*, *64*, 8–15. https://doi.org/10.1016/j.cap.2024.05.011
2. Kim, S. E., Sung, J. Y., Jeon, J. D., Jang, S. Y., Lee, H. M., Moon, S. M., Kang, J. G., Lim, H. J., Jung, H., & Lee, S. W. (2022). Toward advanced high-K and electrode thin films for DRAM capacitors via atomic layer deposition. *Advanced Materials Technologies*, *8*(20). https://doi.org/10.1002/admt.202200878
3. Bersch, E., Rangan, S., Bartynski, R. A., Garfunkel, E., & Vescovo, E. (2008). Band offsets of ultrathin high-κoxide films with Si. *Physical Review B*, *78*(8). https://doi.org/10.1103/physrevb.78.085114
4. Li, S., Liu, X., Yang, H., Zhu, H., & Fang, X. (2024). Two-dimensional perovskite oxide as a photoactive high-κ gate dielectric. *Nature Electronics*, *7*(3), 216–224. https://doi.org/10.1038/s41928-024-01129-9
5. Kingon, A. I., Maria, J., & Streiffer, S. K. (2000). Alternative dielectrics to silicon dioxide for memory and logic devices. *Nature*, *406*(6799), 1032–1038. https://doi.org/10.1038/35023243
6. Kong, X., Yang, L., Meng, F., Zhang, T., Zhang, H., Lin, Y., Huang, H., Zhang, S., Guo, J., & Nan, C. (2025). High-entropy engineered BaTiO3-based ceramic capacitors with greatly enhanced high-temperature energy storage performance. *Nature Communications*, *16*(1), 885. https://doi.org/10.1038/s41467-025-56195-0

7. Li, D., Liu, Z., Zhao, W., Guo, Y., Wang, Z., Xu, D., Huang, H., Pang, L., Zhou, T., Liu, W., & Zhou, D. (2025). Global-optimized energy storage performance in multilayer ferroelectric ceramic capacitors. *Nature Communications*, *16*(1), 188. https://doi.org/10.1038/s41467-024-55491-5
8. An, J., Ahn, J., Lim, Y., Bae, H. B., Ryu, J., & Chung, S. (2025). Microstructure optimization via Grain-Boundary segregation to enhance DC bias dielectric performance of BATIO3 multilayer ceramic capacitors. *Advanced Materials*, e07233. https://doi.org/10.1002/adma.202507233
9. Bahers, T. L., Rérat, M., & Sautet, P. (2014). Semiconductors Used in Photovoltaic and Photocatalytic Devices: Assessing Fundamental Properties from DFT. *The Journal of Physical Chemistry C*, *118*(12), 5997–6008. https://doi.org/10.1021/jp409724c
10. Li, S., Liu, X., Yang, H., Zhu, H., & Fang, X. (2024b). Two-dimensional perovskite oxide as a photoactive high-κ gate dielectric. *Nature Electronics*, *7*(3), 216–224. https://doi.org/10.1038/s41928-024-01129-9
11. Baroni, S., De Gironcoli, S., Corso, A. D., & Giannozzi, P. (2001). Phonons and related crystal properties from density-functional perturbation theory. *Reviews of Modern Physics*, *73*(2), 515–562. https://doi.org/10.1103/revmodphys.73.515
12. Li, Y., Tang, Z., Chen, Z., Sun, M., Zhao, B., Li, H., Tao, H., Yuan, Z., Duan, W., & Xu, Y. (2024). Neural-Network density Functional Theory based on variational energy minimization. *Physical Review Letters*, *133*(7), 076401. https://doi.org/10.1103/physrevlett.133.076401
13. Yi, Y., Wang, L., & Chen, Z. (2021). Adaptive global kernel interval SVR-based machine learning for accelerated dielectric constant prediction of polymer-based dielectric energy storage. *Renewable Energy*, *176*, 81–88. https://doi.org/10.1016/j.renene.2021.05.045
14. Kim, C., Pilania, G., & Ramprasad, R. (2016). Machine learning assisted predictions of intrinsic dielectric breakdown strength of ABX3Perovskites. *The Journal of Physical Chemistry C*, *120*(27), 14575–14580. https://doi.org/10.1021/acs.jpcc.6b05068
15. Ye, Y., Ni, Z., Hu, K., Li, Y., Peng, Y., & Chen, X. (2023c). Dielectric constant prediction of perovskite microwave dielectric ceramics via machine learning. *Materials Today Communications*, *35*, 105733. https://doi.org/10.1016/j.mtcomm.2023.105733
16. Deng, J., & Jia, G. (2022). Dielectric constant prediction of pure organic liquids and their mixtures with water based on interpretable machine learning. *Fluid Phase Equilibria*, *561*, 113545. https://doi.org/10.1016/j.fluid.2022.113545
17. Mao, Z., Li, W., & Tan, J. (2024). Dielectric tensor prediction for inorganic materials using latent information from preferred potential. *Npj Computational Materials*, *10*(1). https://doi.org/10.1038/s41524-024-01450-z
18. Birhane, A., Kasirzadeh, A., Leslie, D., & Wachter, S. (2023). Science in the age of large language models. *Nature Reviews Physics*, *5*(5), 277–280. https://doi.org/10.1038/s42254-023-00581-4
19. Na, Y., Kim, J. J., Park, C., Hwang, J., Kim, C., Lee, H., & Lee, J. (2025). Advanced scientific information mining using LLM-driven approaches in layered cathode materials for sodium-ion batteries. *Materials Advances*, *6*(8), 2543–2548. https://doi.org/10.1039/d5ma00004a
20. Agarwal, S., Mahmood, A., & Ramprasad, R. (2025). Polymer solubility prediction using large language models. *ACS Materials Letters*, *7*(6), 2017–2023. https://doi.org/10.1021/acsmaterialslett.5c00054
21. Antunes, L. M., Butler, K. T., & Grau-Crespo, R. (2024). Crystal structure generation with autoregressive large language modeling. *Nature Communications*, *15*(1), 10570. https://doi.org/10.1038/s41467-024-54639-7
22. Kim, S., Jung, Y., & Schrier, J. (2024). Large language models for inorganic synthesis predictions. *Journal of the American Chemical Society*, *146*(29), 19654–19659. https://doi.org/10.1021/jacs.4c05840
23. Yu, S., Ran, N., & Liu, J. (2024). Large-language models: The game-changers for materials science research. *Artificial Intelligence Chemistry*, *2*(2), 100076. https://doi.org/10.1016/j.aichem.2024.100076
24. Du, Y., Tian, M., Ronanki, S., Rongali, S., Bodapati, S., Galstyan, A., Wells, A., Schwartz, R., Huerta, E. A., & Peng, H. (2025). Context length alone hurts LLM performance despite perfect retrieval. *arXiv*. https://doi.org/10.48550/arxiv.2510.05381
25. Rignanese, G., Gonze, X., & Pasquarello, A. (2001). First-principles study of structural, electronic, dynamical, and dielectric properties of zircon. *Physical Review. B, Condensed Matter*, *63*(10). https://doi.org/10.1103/physrevb.63.104305
26. Ghosez, P., Michenaud, J. P., & Gonze, X. (1998). Dynamical atomic charges: The case ofABO3compounds. *Physical Review. B, Condensed Matter*, *58*(10), 6224–6240. https://doi.org/10.1103/physrevb.58.6224
27. Bersuker, I. B., & Polinger, V. (2020). Perovskite crystals: unique Pseudo-Jahn–Teller origin of ferroelectricity, multiferroicity, permittivity, flexoelectricity, and polar nanoregions. *Condensed Matter*, *5*(4), 68. https://doi.org/10.3390/condmat5040068

28. King, G., Thimmaiah, S., Dwivedi, A., & Woodward, P. M. (2007). Synthesis and characterization of new AA′BWO6 perovskites exhibiting simultaneous ordering of A-Site and B-Site cations. *Chemistry of Materials*, *19*(26), 6451–6458. https://doi.org/10.1021/cm0716708
29. Mittal, P., Chawla, D., Sushant, N., Mehta, J., & Gupta, P. (2025). Perovskite Multiferroics as Energy Harvesters: Exploring the multidimensional properties for energy harvesting applications. In *Materials horizons* (pp. 157–184). https://doi.org/10.1007/978-981-96-7419-0_6
30. Seshadri, R. (2006). Lone pairs in insulating pyrochlores: Ice rules and high-k behavior. *Solid State Sciences*, *8*(3–4), 259–266. https://doi.org/10.1016/j.solidstatesciences.2006.02.020
31. Saha, R. A., Halder, A., Fu, D., Itoh, M., Saha-Dasgupta, T., & Ray, S. (2021). The critical role of stereochemically active lone pair in introducing high temperature ferroelectricity. *Inorganic Chemistry*, *60*(6), 4068–4075. https://doi.org/10.1021/acs.inorgchem.1c00117
32. Bai, W., Zhou, Y., Xu, L., Xiao, H., Tong, Y., He, C., Pang, J., Xie, Q., & Yang, C. (2025). Colossal permittivity and low dielectric loss in (Ta, Li) co-doped SrTiO3 ceramics designed by B-site defect engineering. *Ceramics International*, *51*(29), 61084–61092. https://doi.org/10.1016/j.ceramint.2025.10.304
33. Lu, Y., & Hsiang, H. (2024). Effects of A/B ratio on the dielectric properties and AC conductivity behaviors of La-doped SrTiO3 with colossal permittivity. *Journal of Alloys and Compounds*, *1008*, 176558. https://doi.org/10.1016/j.jallcom.2024.176558
34. Takigawa, A., Kiyohara, S., & Kumagai, Y. (2025b). Accelerated Discovery of High-κ Oxides with Physics-Based Factorized Machine Learning. *arXiv*. https://doi.org/10.48550/arxiv.2509.26022
35. Gopakumar, A., Pal, K., & Wolverton, C. (2022). Identification of high-dielectric constant compounds from statistical design. *Npj Computational Materials*, *8*(1). https://doi.org/10.1038/s41524-022-00832-5
36. Yim, K., Yong, Y., Lee, J., Lee, K., Nahm, H., Yoo, J., Lee, C., Hwang, C. S., & Han, S. (2015). Novel high-κ dielectrics for next-generation electronic devices screened by automated ab initio calculations. *NPG Asia Materials*, *7*(6), e190. https://doi.org/10.1038/am.2015.57
37. Perdew, J. P., & Levy, M. (1983b). Physical content of the exact Kohn-Sham orbital energies: band gaps and derivative discontinuities. *Physical Review Letters*, *51*(20), 1884–1887. https://doi.org/10.1103/physrevlett.51.1884
38. Pallikara, I., Kayastha, P., Skelton, J. M., & Whalley, L. D. (2022). The physical significance of imaginary phonon modes in crystals. *Electronic Structure*, *4*(3), 033002. https://doi.org/10.1088/2516-1075/ac78b3
39. Jain, A., Ong, S. P., Hautier, G., Chen, W., Richards, W. D., Dacek, S., Cholia, S., Gunter, D., Skinner, D., Ceder, G., & Persson, K. A. (2013). Commentary: The Materials Project: A materials genome approach to accelerating materials innovation. *APL Materials*, *1*(1). https://doi.org/10.1063/1.4812323
40. Clark, S. J., Segall, M. D., Pickard, C. J., Hasnip, P. J., Probert, M. I. J., Refson, K., & Payne, M. C. (2005c). First principles methods using CASTEP. *Zeitschrift Für Kristallographie - Crystalline Materials*, *220*(5–6), 567–570. https://doi.org/10.1524/zkri.220.5.567.65075
41. Langreth, D. C., & Perdew, J. P. (1980c). Theory of nonuniform electronic systems. I. Analysis of the gradient approximation and a generalization that works. *Physical Review. B, Condensed Matter*, *21*(12), 5469–5493. https://doi.org/10.1103/physrevb.21.5469
42. Fischer, T. H., & Almlof, J. (1992c). General methods for geometry and wave function optimization. *The Journal of Physical Chemistry*, *96*(24), 9768–9774. https://doi.org/10.1021/j100203a036
43. Khan, A., Banerjee, D., Rawat, D., Nath, T. K., Soni, A., Chatterjee, S., & Taraphder, A. (2025). Emergence of spin–phonon coupling in a Gd-doped Y2CoMnO6 double perovskite oxide: a combined experimental and ab initio study. *Physical Chemistry Chemical Physics*, *27*(34), 18005–18014. https://doi.org/10.1039/d5cp00229j

**Table 1: Precision benchmarking of *DielecMIND* against existing ML-based high-κ dielectric discovery models.** Performance is evaluated using precision, defined as the fraction of predicted candidates that satisfy the target figure-of-merit criteria upon DFT/DFPT verification.

| Modeling Approach | $\Phi_M$ | Validated hits /predicted candidates | Precision (%) | Existence in MP database | Ref |
|---|---|---|---|---|---|
| ML (ANN) | >200 | 1/17 | 5.8 | Yes | 35 |
| **ML (EGNN)** | >300 | 3/60 | 5.0 | Yes | 17 |
| **ML (EGNN + MLP)** | >350 | 38/240 | 15.8 | Yes | 34 |
| **LLM** | >350 | 5/60 | 8.3 | No | This work (Phase II) |
| **Baseline** | >70 | 2/60 | 3.3 | No | This work |

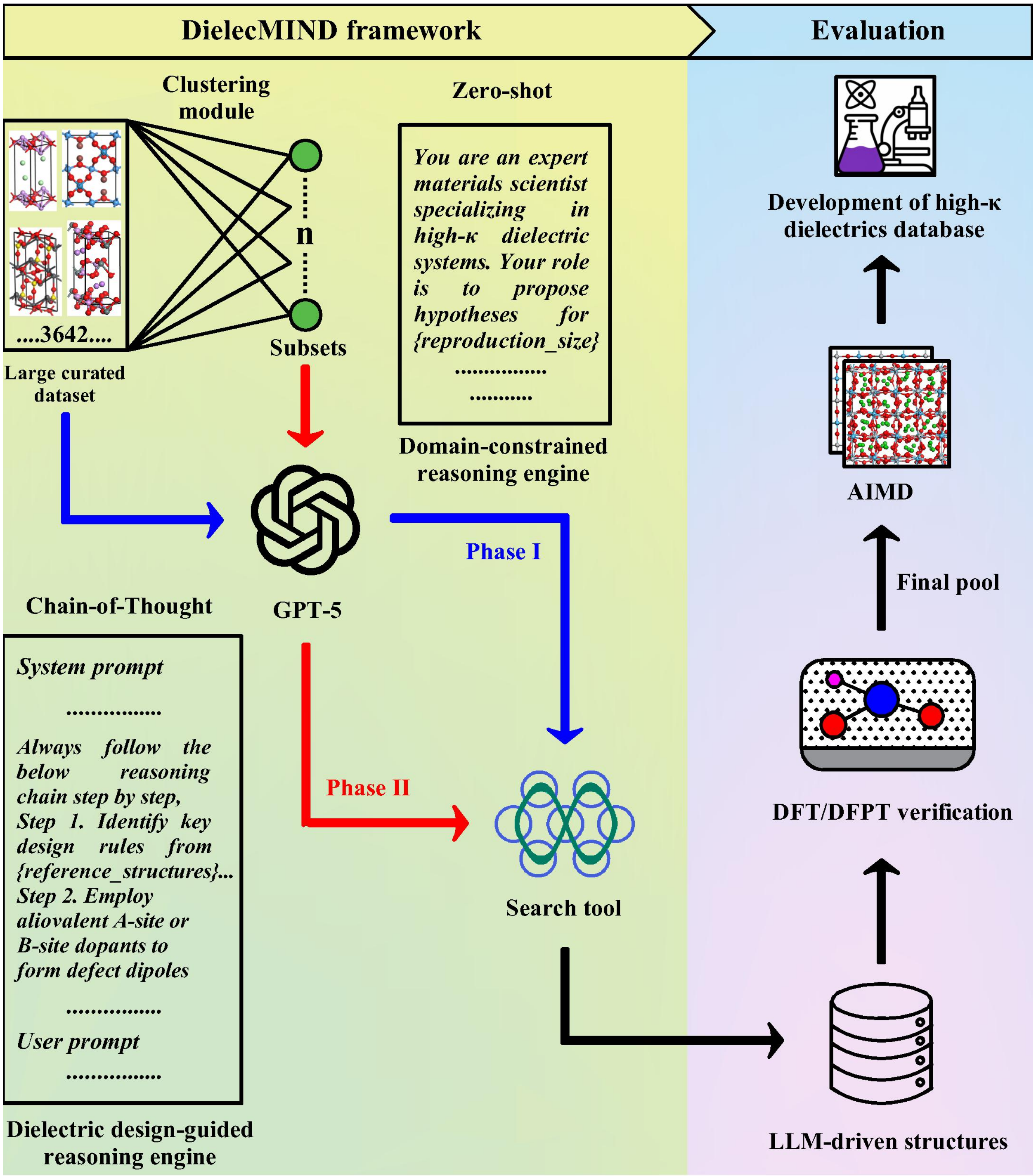

**Fig. 1: Architecture of the *DielecMIND* framework for high-κ dielectric discovery.** *DielecMIND* operates through two synergistic phases. **Phase I** (blue pipeline) employs zero-shot generation, where domain-constrained reasoning engine guides the LLM to propose novel dielectric structures from a large curated dataset. **Phase II** (red pipeline) introduces dielectric material design-guided CoT reasoning, in which clustered reference structures and explicit physical constraints inform iterative hypothesis generation. Across both phases, search tool queries the Materials Project database to validate novelty and filter out previously reported candidates. All LLM-generated materials are rigorously validated through first-principles DFT/DFPT, with thermodynamic stability further assessed via AIMD. This workflow iteratively produces expert-guided, previously unreported high-κ dielectric candidates with quantitative physical verification.

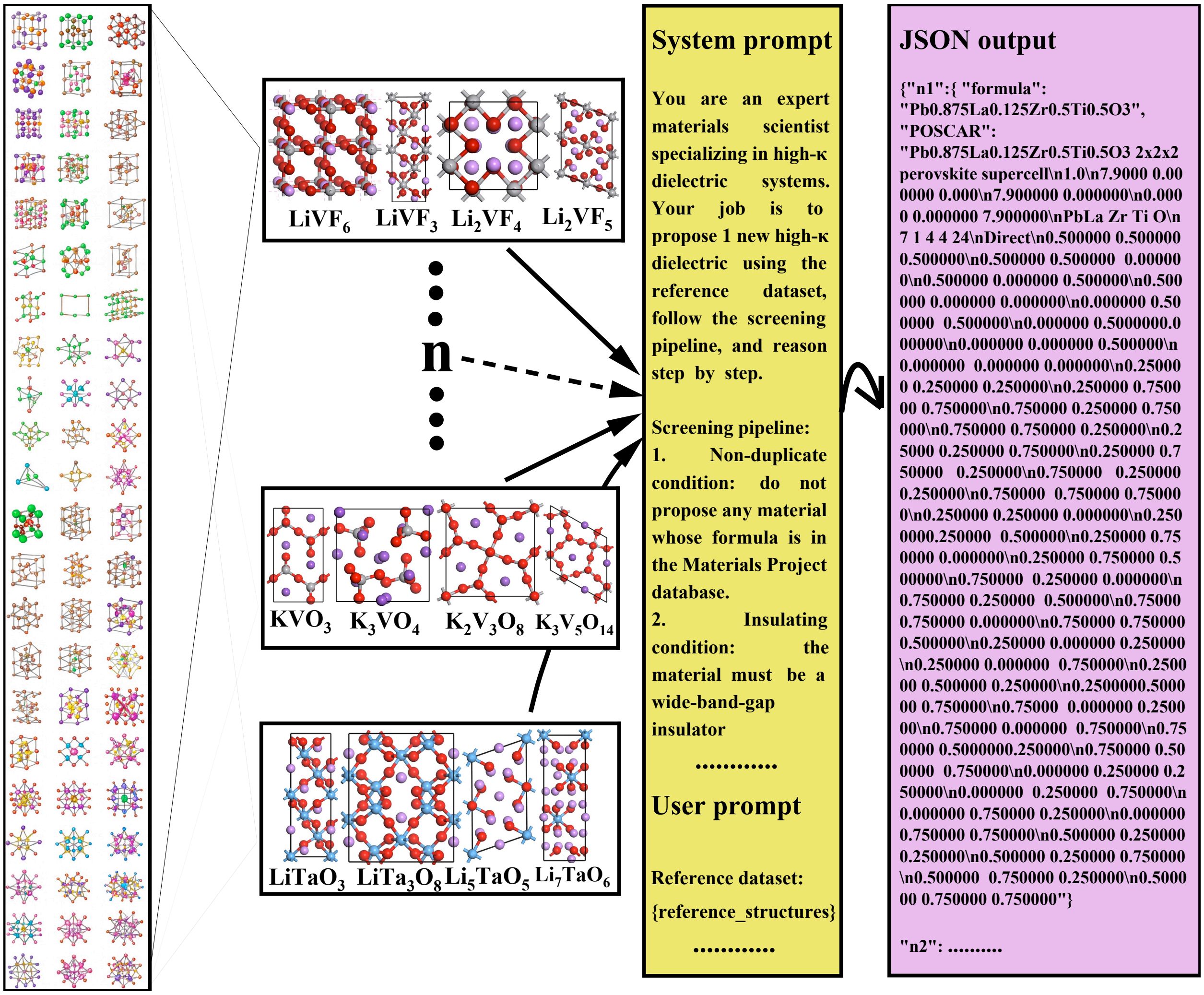

**Fig. 2: Phase II workflow of the DielecMIND framework.** Phase II integrates three coordinated components to enable dielectric material design-guided, chain-of-thought reasoning over the large curated dataset. A clustering module first partitions the large curated dataset into chemically similar subsets, for example [Li, V, F] → {$LiVF_6$, $LiVF_3$, $Li_2VF_4$, $Li_2VF_5$}, [K, V, O] → {$KVO_3$, $K_3VO_4$, $K_2V_3O_8$, $K_3V_5O_{14}$}, and [Li, Ta, O] → {$LiTaO_3$, $LiTa_3O_8$, $Li_5TaO_5$, $Li_7TaO_6$}. Each subset is then processed by a dielectric design-guided reasoning engine that performs step-by-step reasoning using lattice-level dielectric design principles to propose new candidate compositions. A search tool subsequently verifies structural novelty

against the Materials Project database before validated candidates are serialized in JSON format for first-principles evaluation.

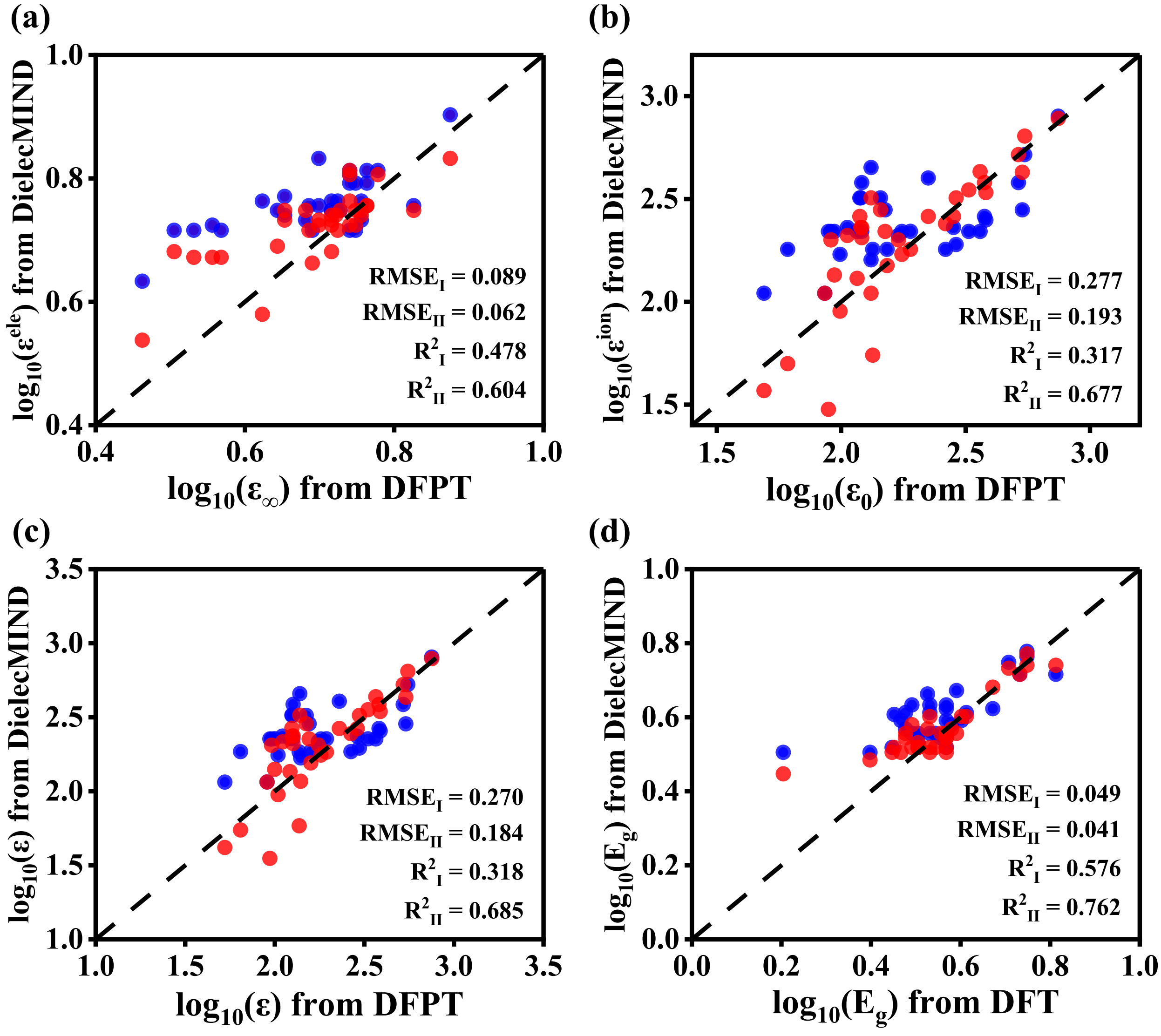

**Fig. 3: Performance comparison between *DielecMIND* predictions and first-principles reference data.** Parity plots compare *DielecMIND*-predicted values against literature-sourced DFT/DFPT ground truth for 34 benchmark materials on a base-10 logarithmic scale. Blue circles denote Phase I predictions, while red circles correspond to Phase II predictions. Shown are comparisons for (a) electronic dielectric constant $\varepsilon_\infty$, (b) ionic dielectric constant $\varepsilon_0$, (c) total dielectric constant $\varepsilon$, and (d) band gap $E_g$.

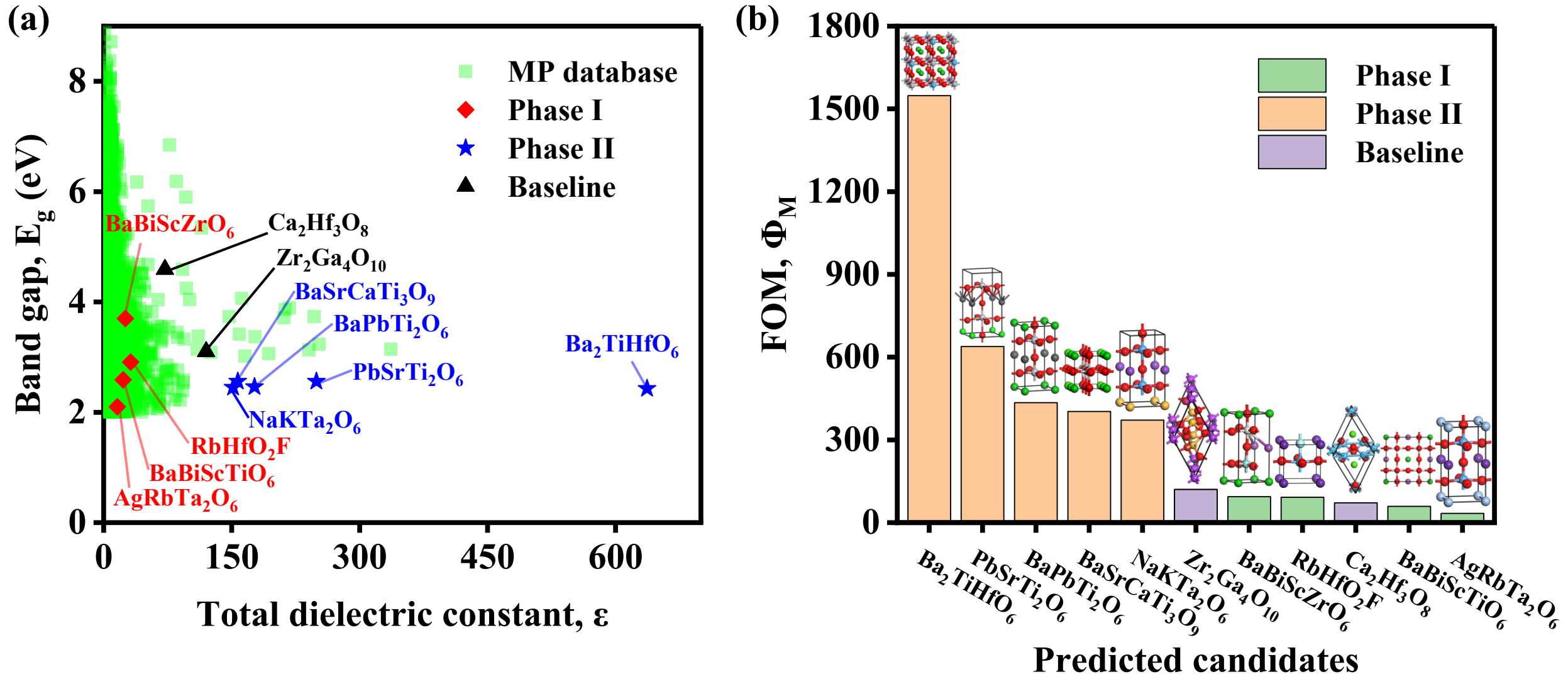

**Fig. 4: *DielecMIND*-predicted and DFT/DFPT-validated high-*κ* dielectrics.** (a) Dielectric constant (ε) plotted against the band gap ($E_g$) for the nine candidate dielectric materials identified by DielecMIND in two distinct phases, shown alongside the curated Materials Project dataset for comparison. Two dielectric materials identified through the baseline response. (b) Figure of merit (FOM) of the identified high-*κ* dielectrics, annotated with their corresponding crystal structures, highlighting materials that simultaneously combine large dielectric response and wide band gap. The 5 newly discovered high-*κ* materials increase the total number of known $\kappa > 150$ dielectrics by ≃35%, underscoring the statistical significance of each validated discovery in this sparsely populated regime.

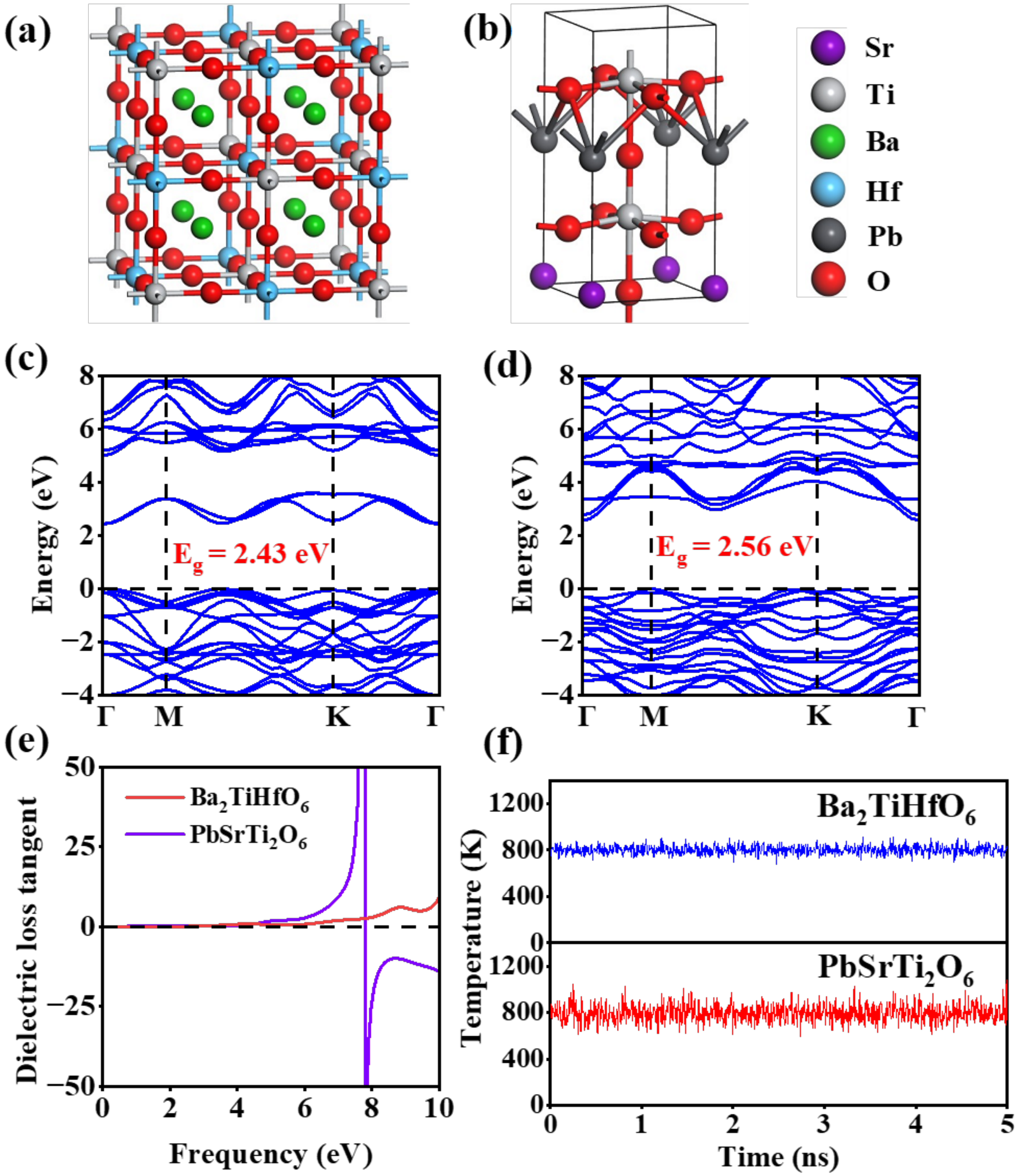

**Fig. 5: Theoretical characterization of the two highest-ranked dielectric candidates.** (a,b) Crystal structure representations of $Ba_2TiHfO_6$ and $PbSrTi_2O_6$, respectively. Purple, silver, green, cyan, dark gray, and red spheres denote Sr, Ti, Ba, Hf, Pb, and O atoms. (c,d) Electronic band structures of $Ba_2TiHfO_6$ and $PbSrTi_2O_6$, confirming their insulating character. (e) Frequency-dependent dielectric loss tangent for both compounds. (f) AIMD simulations demonstrating the thermal stability of $Ba_2TiHfO_6$ and $PbSrTi_2O_6$.